\documentclass[aps, prd, twocolumn, superscriptaddress, nofootinbib]{revtex4}
\usepackage{graphicx}% Include figure files
\usepackage{dcolumn}% Align table columns on decimal point
\usepackage{amssymb}% outlined letters
\usepackage{amsmath,amssymb,amsfonts}

\begin{document}

%Macros
\newcommand{\Eq}[1]{\mbox{Eq. (\ref{eqn:#1})}}
\newcommand{\Fig}[1]{\mbox{Fig. \ref{fig:#1}}}
\newcommand{\Sec}[1]{\mbox{Sec. \ref{sec:#1}}}

\newcommand{\PHI}{\phi}
\newcommand{\PhiN}{\Phi^{\mathrm{N}}}
\newcommand{\vect}[1]{\mathbf{#1}}
\newcommand{\Del}{\nabla}
\newcommand{\unit}[1]{\;\mathrm{#1}}
\newcommand{\x}{\vect{x}}
\newcommand{\y}{\vect{y}}
\newcommand{\p}{\vect{p}}
\newcommand{\ScS}{\scriptstyle}
\newcommand{\ScScS}{\scriptscriptstyle}
\newcommand{\xplus}[1]{\vect{x}\!\ScScS{+}\!\ScS\vect{#1}}
\newcommand{\xminus}[1]{\vect{x}\!\ScScS{-}\!\ScS\vect{#1}}
\newcommand{\diff}{\mathrm{d}}
\newcommand{\mk}{{\mathbf k}}
\newcommand{\ep}{\epsilon}

\newcommand{\be}{\begin{equation}}
\newcommand{\ee}{\end{equation}}
\newcommand{\bea}{\begin{eqnarray}}
\newcommand{\eea}{\end{eqnarray}}
\newcommand{\vu}{{\mathbf u}}
\newcommand{\ve}{{\mathbf e}}
\newcommand{\vn}{{\mathbf n}}
\newcommand{\vk}{{\mathbf k}}
\newcommand{\vz}{{\mathbf z}}
\newcommand{\vx}{{\mathbf x}}
\def\dup{\;\raise1.0pt\hbox{$'$}\hskip-6pt\partial\;}
\def\ddn{\;\overline{\raise1.0pt\hbox{$'$}\hskip-6pt\partial}\;}

%=====================================================================
%=====================================================================
%=====================================================================

\title{The phenomenology of squeezing and its status in non-inflationary theories}

\newcommand{\addressImperial}{Theoretical Physics, Blackett Laboratory, Imperial College, London, SW7 2BZ, United Kingdom}
\newcommand{\addressRoma}{Dipartimento di Fisica, Universit\`a ‚ÄúLa Sapienza‚Äù
and Sez. Roma1 INFN, P.le A. Moro 2, 00185 Roma, Italia}

\author{Giulia Gubitosi}
\affiliation{\addressImperial}
\author{Jo\~{a}o Magueijo}
\affiliation{\addressImperial}

\date{\today}

\begin{abstract}
In this paper we skim the true phenomenological requirements behind the concept of inflationary squeezing.
We argue that all that is required is that at horizon re-entry the fluctuations form standing waves with the correct temporal phase
(specifically, sine waves). We quantify this requirement and relate it to the initial conditions fed into the radiation dominated
epoch by whatever phase of the Universe produced the fluctuations. The only relevant quantity turns out to be the degree of
suppression of the momentum, $p$,
of the fluctuations, $y$, which we measure by $\sigma\sim \omega^2 |y|^2/|p|^2$. Even though $\sigma$ equals 
the squeezing parameter, $s$, in the case of inflation and bimetric varying speed of light scenarios, this is not true in general, 
specifically 
in some bouncing Universe models. It is also not necessary to produce a large $\sigma$ at the end of the primordial phase: it is enough that  $\sigma$ be not too small.
This is the case with scenarios based on modified dispersion relations (MDR) emulating the dispersion relations of Horava-Lifshitz theory, which produce $\sigma\sim 1$, enough to comply with the observational requirements. Scenarios based on MDR leading to a slightly red spectrum are also examined, and shown to satisfy the observational constraints. 
\end{abstract}


\maketitle

\section{Introduction}

Squeezing of quantum fluctuations has for a long time been heralded as a key feature of inflation, starting with its forefathers (e.g.~\cite{grish, Grishchuk:1993zd,starob96}). As the vacuum fluctuations leave the horizon their dynamics undergoes a transition from harmonic to ``inverted" harmonic oscillator. A Gaussian  wavefuction with equal spread in position and momentum will then ``squeeze'' along a straight line in phase space. This imposes a perfect correlation between position and momentum; hence the claim that quantum uncertainty is replaced by classical statistics. Squeezing has even 
been blamed for the transition to  classicality of quantum generated fluctuations, in a process known as ``decoherence without decoherence''~\cite{starob96}. This was disputed variously~\cite{pedro,martin, Martin:2012pea}, and it now seems that 
classicality requires an adaptation of the usual decoherence mechanisms~\cite{Kiefer:1998qe, Kiefer:2008ku, Perez:2005gh,Nelson:2016kjm,Burgess:2014eoa}.

It is easy to argue that the basic process of ``squeezing'' has nothing to do with quantum mechanics (although, of course, it interacts with it).  We can imagine a scenario where the wave-function collapse took place before first horizon crossing (e.g.~\cite{drubo}), so that, whatever squeezing does, it must apply to quantum fluctuations already turned classical. We can also examine the effects of ``squeezing'' in scenarios where the fluctuations are thermal in nature, or more generally stochastic classical waves from the start. Then, whatever ``squeezing'' stands for, it must be understandable classically. The crucial issue is: what is the phenomenological content of ``squeezing'' that stands as a requirement for all theories, beyond its quantum luggage?

We start answering this question in Section~\ref{phen}, where we contrast how QFT sets up Fourier analysis (with two independent traveling waves for each headless direction), and the Fourier transform used in astrophysical cosmology and late time codes for evaluating observables (such as Boltzmann codes). The latter has half the number of degrees of freedom of the former.  We trace this to the phenomenological requirement that any waves at late times must be standing waves, built from travelling waves moving along opposite directions with equal real amplitude \cite{grish}. The nodes of the standing waves are shifted at random within the ensemble (so that translational invariance is preserved); however, their time-phase must be fixed on horizon re-entry. Specifically the time function must match the phase of a sine wave, in order to comply with CMB phenomenology, namely the phase of the Doppler peaks. We quantify this requirement with a quantity $\Sigma$ measuring the strength of the sine wave with respect to the cosine wave. 

This must be the endpoint of ``squeezing'', now standing for whatever happens to fluctuations while they are outside the horizon, waiting to re-enter. However, when 
we examine the initial conditions fed into the radiation dominated Universe (Section~\ref{connector}), we find that the only relevant quantity is $\sigma$, measuring the relative strength of the fluctuations' ``position'' mode over the ``momentum'' mode. After reviewing the squeezing framework (Section~\ref{squeeze}), 
we show (Section~\ref{dSinflation}) that $\sigma$ equals the squeezing parameter $s$ in the inflationary Universe. However this is not true in general, as exemplified by a
contracting radiation dominated Universe, where $\sigma$ and $s$ have opposite behaviours (Section~\ref{contract}). Also, the large $s=\sigma$ injected into the radiation epoch by the inflationary Universe is not strictly needed, and in fact is ``overkill''. This is because in the radiation dominated phase the momentum mode is a decaying mode, so it suffices that the $\sigma$ provided by the previous phase be {\it not too small}. Only an abnormally low $\sigma$ could lead to a sizeable cosine mode at re-entry, and so a travelling wave on re-entry, or a standing wave with the wrong temporal phase. This is shown both in Section~\ref{connector}, in general, and in Section~\ref{rad-reentry}, using the squeezed state formalism. 

The rest of the paper is spent examining the status of alternative models with regards to the production of standing waves in the late Universe.
Bimetric VSL models are found to be almost identical with inflation (Section~\ref{VSL}), and produce $\sigma\gg 1$, outdoing the actual requirement. 
Models based on Horava-Lishitz deformed dispersion relations are very different, and like string gas cosmology, produce $\sigma\sim 1$ at the start of the standard radiation epoch, sufficient to satisfy the phenomenological requirements at horizon re-entry (Section~\ref{DSR}). Scenarios based on DSR leading to a slightly red spectrum are examined in Section~\ref{DSRred}.

In a concluding Section we provide some intuition regarding our results and a summary of the implications for the various scenarios.

\section{The general phenomenological content of ``squeezing''}\label{phen}
In this Section we argue that ``squeezing'' does nothing but impose two important phenomenological requirements. Firstly, it halves the number of Fourier modes for a given headless vector. This results from the imposition of a constraint between the fluctuations curvature variable $y$
and its conjugate momentum\footnote{In this paper we shall used the conventions of~\cite{martin} and thus call $y$ to what in~\cite{Mukhanov:1990me} is called $v$.}.
It forces any wave after horizon re-entry to be a standing wave, i.e. the superposition of 
travelling modes moving in opposite directions with the same (real) amplitude. Secondly, 
the {\it temporal} phase of these standing waves is fixed: they must be sine waves. This ensures that the  position of the Doppler peaks in the CMB spectrum matches the observed one.

\subsection{Halving of the modes}
There seems to be a dichotomy in the way Fourier analysis is carried out in QFT and in observational cosmology. In late-time cosmology one writes \cite{Mukhanov:1990me, martin}:
\be\label{FT1}
y(\vx,\eta)=\int \frac{d\vk}{(2\pi)^3} \, e^{i\vk\cdot \vx}y(\vk,\eta),
\ee
with $y(-\vk,\eta)=y^\star (\vk,\eta)$ enforcing the reality of $y(\vx,\eta)$. This is done in observational cosmology but also in any theoretical tool for computing late time observables, such as Boltzmann codes, where the initial conditions appear 
as the only relics from the early universe.

Instead, in flat space QFT (and in setting up initial conditions before first Hubble crossing), one writes:
\be\label{QFT}
y(\vx,\eta)=\int  \frac{d\vk}{(2\pi)^3\sqrt {2\omega}} \, \left[c(\vk)e^{-i k_\mu x^\mu} + c^\dagger (\vk)e^{i k_\mu  x^\mu}\right]
\ee
where the $c(\vk)$ can be classical amplitudes or quantum (creation and annihilation) operators, depending on whether we consider the expression before or after quantization. Their label $\vk$ points to the direction of motion of the travelling wave, so that $c(\vk)$ and $c(-\vk)$ are independent. The reality condition is automatically enforced for each of the two modes, separately. 

This is not often pointed out,
 but the second expression has twice the number of degrees of freedom of the first. 
This is true even before quantization, so the doubling cannot be blamed on quantum theory. Specifically, in Minkowski space-time, we have:
\bea
y(\vk,\eta)&=&\frac{c(\vk,\eta)+c^\dagger (-\vk,\eta)}{\sqrt{2\omega }} \\
&=&\frac{e^{-i\omega \eta}}{\sqrt{2\omega}}c(\vk)+
\frac{e^{i\omega\eta}}{\sqrt{2\omega}}c^\dagger (-\vk)\label{ywave}
\eea
(with $\omega=c_s k$, where $c_s$ is the speed of propagation).
We see that one complex degree of freedom (viz. $y(\vk,\eta)$) in Eq.~(\ref{FT1}) is split into  two complex degrees of freedom 
(viz. $c(\vk,\eta)$ and $c(-\vk,\eta)$) in Eq.~(\ref{QFT}).  

How does this late time reduction take place? The blanket term usually used for the explanation is the ``squeezing'' of modes that took place while they were outside the horizon. Later in this paper we will show with examples that ``squeezing'' is a just fancy word for loss of conjugate momentum, enforced by a number of reasons, depending on the scenario. All that happens is that for all practical purposes while the modes are outside the horizon we enforce a constraint tying $y$ to its conjugate momentum\cite{Mukhanov:1990me, martin}:
\be\label{pexp}
p=y'-\frac{z'}{z}y
\ee
where $z=a/c_s$ in almost all models (the exception being MDR models, for which $z=a$; see~\cite{Magueijo:2008yk}).
Since~\cite{martin}:
\bea
y(\vk,\eta)&=&\frac{c(\vk,\eta)+c^\dagger (-\vk,\eta)}{\sqrt{2\omega}} \label{eq:yvsc}\\
p(\vk,\eta)&=&-i \sqrt{\frac{\omega}{2}}(c(\vk,\eta)- c^\dagger (-\vk,\eta))\label{eq:pvsc}
\eea
this constraint is equivalent to a constraint between $c(\vk)$ and $c(-\vk)$, which is set as an initial condition in the late time universe.

To phrase the discussion in another way, one should compare the number of degrees of freedom encoded in the pair of conjugate variables $\{y(\vk),p(\vk)\}$ and in $\{c(\vk),c(-\vk)\}$. By complementing (\ref{FT1}) with $p$, a priori its number of degrees of freedom is the same as in (\ref{QFT}). ``Squeezing'' introduces a constraint rendering the momentum a dependent variable on $y$, so that $p$ goes from being statistically uncorrelated with $y$ to being perfectly correlated with it. This is equivalent to introducing a constraint between $c (\vk)$ and $c(-\vk)$, and is behind the reduction in the number of modes. It forces any late time waves to be standing waves, as we now see.

\subsection{Standing waves and their phases}
A standing wave is one that can be factored into a real time-oscillating function and a real spatial-oscillating function:
\be\label{standing}
y_{st.}=A(\overline \vk)\cos(\omega\eta +\phi_t)\cos(\vk\cdot \vx +\phi_x)
\ee
(here $\overline\vk$ stands for a headless vector, since the direction of $\vk$ does not matter). 
As such the wave has fixed nodes in space, and so carries no spatial momentum $P_i$. It is important not  to confuse the spatial momentum of the wave, $P_i$, (which is the Noether charge associated with translational invariance), with the momentum, $p$, defined as the conjugate variable to $y$. The $p$ of a standing wave is not identically zero, but $P_i$ is. 

Expression (\ref{standing}) 
is to be distinguished from the one valid for travelling waves, where the {\it real} time and space oscillations cannot be factored:
\be\label{travelling}
y_{tr.}=B( \vk)\cos(-\omega\eta+\vk\cdot \vx +\phi)\,.
\ee
Although the two expressions should not be confused, 
a generic standing wave can be built from two travelling waves of equal real amplitude moving in opposite directions:
\bea
y&=&B(\vk) \cos(-\omega\eta + \vk\cdot \vx +\phi_+) \nonumber\\
&+&B(-\vk) \cos(\omega\eta + \vk\cdot \vx +\phi_-) \,.\label{twotravelling}
\eea
Setting $B(\vk)=B(-\vk)=B(\overline\vk)$  we obtain:
\be
y =\frac{B(\overline\vk)  }{2} \left(e^{i(-\omega\eta +\phi_+)}+ e^{i(\omega\eta +\phi_-)}\right)e^{i\vk\cdot \vx} + c.c.
\ee
After some algebra we  find that this is a wave of the form (\ref{standing}) with $A=2B$ and 
\bea
\phi_t&=&\frac{\phi_- - \phi_+}{2}\label{phit}\\
\phi_x&=&\frac{\phi_++\phi_-}{2}.\label{phix}
\eea
(the algebra can be simplified by performing a shift back and forth to $\eta'= \eta - \frac{\phi_+-\phi_-}{2\omega}$ so as to equalise the phases of the two modes).

We see that the 4 degrees of freedom of the two unconstrained travelling waves moving along a headless direction become the 3 degrees of freedom of a standing wave, upon the constraint that their real amplitude is the same. The real amplitude of the waves is equated but the phases, although not changing in number, change in character. A priori, one should distinguish between the spatial phase and the temporal phase; however for a traveling wave the two concepts are intertwined, and for each headless vector we must independently randomize the phase of the two waves moving in opposite directions in order to preserve translational invariance. For standing waves the concepts are separate and translational invariance is enforced by randomizing the phase of the spatial wave alone, thereby randomizing the position of its nodes. 
The temporal phase does not need to be randomized to preserve translational invariance. We shall expand on this matter elsewhere~\cite{prep1}.

We now connect these remarks to the late time cosmological framework. 
Let us give $\infty$ subscripts to the amplitudes of expression  (\ref{ywave}), to denote that it refers to a late time expansion (and distinguish it from the early time initial conditions). Thus,
 \be\label{solrad0}
y(\vk)=\frac{e^{-i\omega \eta}}{\sqrt{2\omega}} c_\infty(\vk)+
\frac{e^{i\omega \eta}}{\sqrt{2\omega }}c_\infty^\dagger (-\vk). 
\ee
Inserting this expression into (\ref{FT1}) and adding the complex conjugate (coming from $y(-\vk)$), one finds the two travelling waves moving in 
opposite directions: 
\bea
y&=&e^{i\vk\cdot \vx}y(\vk)+e^{-i \vk\cdot\vx}y(-\vk)\nonumber\\
&=&B(\vk) \cos(-\omega\eta + \vk\cdot \vx +\phi_+) \nonumber\\
&+&B(-\vk) \cos(\omega\eta + \vk\cdot \vx +\phi_-) \label{travelling1}
\eea
where the real amplitudes $B(\vk)$ and $B(-\vk)$ and phases $\phi_+$ and $\phi_-$ are related to $c_\infty(\vk)$ and $c_\infty(-\vk)$
according to:
\be
\frac{c_\infty(\pm \vk)}{\sqrt{2\omega}}=\frac{B(\pm \vk)}{2}e^{\pm i\phi_\pm}.
\ee
Production of late time standing waves requires that $|c_\infty(\vk)|=|c_\infty(-\vk)|$. Their phases, in turn, fix $\phi_x$ and $\phi_t$. The spatial phase $\phi_x$ should be random, so $c_\infty(\vk)$ and $c_\infty(-\vk)$ should be both multiplied an overall random phase (see~(\ref{phix})). The difference of their phases, however, gives $\phi_t$ and this is heavily constrained by observations. 

Specifically, using coordinates where $\eta\rightarrow 0$ at the Big Bang (or at the end of whatever phase generated the fluctuations) the temporal wave of the standing wave must be a sine wave. Otherwise the position of the Doppler peaks would come out wrong, or indeed the peaks themselves could be erased if the temporal phase were random, as is the case with defects and other active incoherent fluctuations~\cite{andyprl}. 
In order to have a temporal sine wave, we must set the temporal phase in (\ref{standing})
\be\label{temp-phase}
\phi_t=\pm \frac{\pi}{2}.
\ee
Effectively this requires  $c_\infty(\vk)= - c_\infty^\dagger (-\vk)$ for any late time waves, so that:
\be
\phi_+=\phi_- \pm \pi \label{squeezedphases}
\ee
producing a late time standing wave with the correct temporal phase. 

\subsection{A measure suited to the phenomenological requirement}\label{SigmaSec}
We now quantify the phenomenological requirement laid down in the last subsection. We start by noting that
although a standing wave with general time phase has fewer degrees of freedom (viz, 3) than the most general solution (which has 4), it is possible to span the most general solution from a  superposition of sine and cosine standing waves.
Specifically, 
the most general solution can be written either as two travelling waves, as in (\ref{solrad0}), or as the two standing waves:
\be\label{solrad1}
y(\vk)= S_1\cos(\omega\eta)+S_2\sin(\omega\eta)
\ee
since equating the two expressions gives:
\bea
S_1&=&\frac{c_\infty(\vk)+c_\infty^\dagger(-\vk)}{\sqrt{2\omega}}\\
S_2&=&\frac{c_\infty(\vk)-c_\infty^\dagger(-\vk)}{\sqrt{2\omega}i}.
\eea
Obviously the $S_i$ (with $i=1,2$) have to be complex in general (with $S_i=\rho_ie^{\phi_i}$)
implying separate spatial phases for the two waves. Indeed when (\ref{solrad1}) is inserted in 
(\ref{FT1}) and the complex conjugate is added we obtain:
\bea
y&=&2\rho_1\cos(\omega\eta)\cos(\vk \cdot \vx +\phi_1)\nonumber \\
&+&2\rho_2\sin(\omega\eta)\cos(\vk \cdot \vx +\phi_2)\label{ystanwave}
\eea
Thus, time sine and cosine standing waves (if allowed independent spatial phases) are just
a basis for the most general solution, as much as the usual travelling wave basis. 

The standing wave basis is best suited for the definition of 
a quantity that can be constrained experimentally, in view of the last subsection. Given that the most general late time solution can be
written as (\ref{solrad1}) we can define in general:
\be
\Sigma=\left|\frac{S_2}{S_1}\right|^2.
\ee
Then $\Sigma\gg 1$ stands for the requirement that at late times a  standing wave of the right temporal  phase be produced. The usual constraint on isocurvature modes can be used to constrain $\Sigma$ directly.

\section{The connector with the Early Universe}\label{connector}
Given that all that is observationally needed is  $\Sigma\gg 1$, one may wonder what is actually required from the phase that produced
the primordial fluctuations in order to ensure it, and for which inflationary squeezing is often blamed. Whatever this is, it  must be inserted as an initial condition
into an expanding radiation dominated epoch, from the previous phase that produced the fluctuations. In gluing the fluctuations from the two phases 
one must match $\zeta=y/z$ and $\zeta'=p/z$ (in practice, in this paper, this will be the same as matching $y$ and $p$, since only their ratio is relevant). In general there will be two modes, one with $y$ but no $p$, another with $p$ but no $y$. 
One should therefore define a quantity measuring the relative strength of the two modes at the end of the primordial phase.

We propose:
\be\label{sigma}
\sigma(\vk) = \frac{| y(\vk)|^2\omega^2 }{4|p(\vk)|^2}
\ee
for  a measure of the relative strength of the two modes. We will show later that this quantity reduces to the ``squeezing parameter'' 
in the case of inflation. However it is more general, and it has the virtue of connecting directly to observable $\Sigma$, as we now show.

The gluing is invariably done with  $\omega \eta\ll 1$ in the expanding radiation phase (with  $\eta>0$), and in this limit
(\ref{solrad1}) produces:
\bea
y&\approx& S_1 +\ S_2 \omega \eta\label{yap}\\
p&\approx&-\frac{S_1}{\eta}.\label{pap}
\eea
By matching $y$ and $p$ from a previous phase we have to match the value of $\sigma$ at the end of that phase to  the value $\sigma_{0}$ at the beginning of the expanding radiation phase:
\be
\sigma_0\approx \frac{(\omega\eta)^2}{4}{\left|1+\frac{S_2}{S_1}\omega\eta\right|}^2,
\ee
obtained by inserting (\ref{yap}) and (\ref{pap}) into (\ref{sigma}).
We see that we do not need to require $\sigma=\sigma_0\gg 1$ to ensure $\Sigma\gg 1$. As long as $\sigma=\sigma_0$ 
is not much smaller than 1 we have:
\be\label{Sigma-sigma}
\Sigma\approx \frac{4\sigma_0}{(\omega\eta_0)^4}\gg 1.
\ee
The reason is that
the momentum mode, capable of generating a time cosine standing wave, is a decaying mode in the expanding radiation phase. Its survival at re-entry would require a very large
initial condition. As long as the momentum mode is comparable (or, in fact, simply not ridiculously larger) than the momentum-free
mode, only the latter survives, and it produces a time sine standing wave.

Although we have illustrated this point with a pure radiation Universe, introducing a matter epoch does not qualitatively change the discussion. Indeed the discussion is valid for any equation of state, although the discussion is more complicated.  The two basis described in Section~\ref{SigmaSec} are then Bessel functions (for standing waves) and Hankel functions (for travelling waves). In general they may be mapped into one another; however only one of the Bessel modes is regular at the origin (the growing mode), so natural boundary conditions select out the other Bessel component. This forms a standing wave with the correct temporal phase.

\section{The formalism of squeezing}\label{squeeze}
We now review the formalism of squeezing (following the notation of~\cite{martin}), stressing that, although it may always be used, it is neither necessary, nor does its quantum interpretation need to be imported into the discussion.  In addition, as we shall see in several examples later in this paper, the formalism should be used with caution.

The basic idea is that general solutions for $\{y(\vk,\eta),p(\vk,\eta)\}$  may be expressed in the basis defined by the initial conditions, and these 
may be expressed as travelling waves labelled by $c_0(\vk)$ and $c^\dagger_0(-\vk)$ (identified from (\ref{ywave})). 
Thus, 
\bea
y(\vk,\eta)&=&f_\vk(\eta)c_0(\vk)+  f^\star _\vk(\eta)c^\dagger _0(-\vk) \label{eq:yvsc0}\\
p(\vk,\eta)&=&-i[g_\vk(\eta)c_0(\vk)-  g^\star _\vk(\eta)c^\dagger _0(-\vk)]\label{eq:pvsc0}
\eea
with $g=i(f'-\frac{z'}{z}f)$. When we insert this into (\ref{eq:yvsc}) and (\ref{eq:pvsc}) we see that 
the two functions  $\{c(\vk,\eta),c^\dagger (-\vk,\eta)\}$ do not need to align with $c_0(\vk)$ and $c^\dagger_0(-\vk)$, 
as they do in (\ref{ywave}) (valid at all times for Minkowski space-time only). In general
\bea
c(\vk,\eta)&=&u_\vk(\eta)c_0(\vk)+  v_\vk(\eta)c^\dagger _0(-\vk) \label{eq:cEvolution}\\
c^\dagger (-\vk,\eta)&=&v^\star_\vk(\eta)c_0(\vk) + u^\star _\vk(\eta)c^\dagger _0(-\vk)\label{eq:cdagEvolution}
\eea
with
\bea
u_\vk &=&\frac{1}{2}\left(\sqrt{2\omega} f_\vk  + \sqrt{\frac{2}{\omega}} g_\vk\right) \label{uvsfg}\\
v_\vk^\star &=&\frac{1}{2}\left(\sqrt{2\omega} f_\vk  - \sqrt{\frac{2}{\omega}} g_\vk\right),\label{vvsfg}
\eea
or
\bea
f_\vk&=&\frac{u_\vk+v_\vk^\star}{\sqrt{2\omega}}\label{fexp}\\
g_\vk&=&\sqrt{\frac{\omega}{2}}(u_\vk-v_\vk^\star)\label{gexp}.
\eea
Equations (\ref{eq:cEvolution}) and (\ref{eq:cdagEvolution}) form a
Bogolubov transformation, and requiring that the evolution of $c(\vk,\eta)$ and $c^{\dagger}(-\vk,\eta)$ is unitary imposes the constraint 
\be\label{unit-condition}
|u_{k}|^2 -|v_{k}|^2=1
\ee
(this can also be derived simply from the Wronskian condition). 
Therefore, we can  parametrize:
\bea
u_{k}(\eta)&=&e^{-i\theta_{k}(\eta)} \cosh(r_{k}(\eta))\label{uexp}\\
v_{k}(\eta)&=&e^{i(\theta_{k}(\eta)+2 \phi_{k}(\eta))} \sinh(r_{k}(\eta))\label{vexp}\, .
\eea
Squeezing is defined as a condition of maximal mixing (or ``particle production''), obtained via
the squeezing parameter $s$:
\be
s_\vk(\eta)=|v_\vk(\eta)|^2\gg 1. 
\ee
From (\ref{unit-condition}) we see that this implies $|u|^2\approx  |v|^2\gg 1$. Squeezing may also be quantified by the ``angle'' $r_k$, and identified from  $r_{k}\gg 1$.

As an indication, we can see that for ``generic initial conditions'', when $r_k\rightarrow \infty$ we can write very roughly:
\be
\sigma\sim \frac{|1+e^{-2i\phi}|^2}{|1-e^{-2i\phi}|^2},
\ee
(this is obtained by inserting (\ref{fexp}) and (\ref{gexp})  into (\ref{eq:yvsc0}) and (\ref{eq:pvsc0}), and then the latter into (\ref{sigma}), while being cavalier about the dependence on the initial conditions $c_0(\vk)$). 
Therefore the relation between squeezing and $\sigma\gg 1$ is at best dependent on the squeezing angle $\phi$, and we can have $\sigma\gg1$, $\sigma\ll 1$ or $\sigma\sim 1$ associated with squeezing, something we shall highlight with examples in the next Section. In fact, looking at the detailed expression obtained, we see that in general the value of $\sigma$ also depends on $\theta$ and the initial conditions $c_0(\vk)$. As we shall see in the next Section, for the specific values of $\theta$ and $\phi$ found in topical cases it just so happens that the dependence on the initial conditions drops out of $\sigma$.

\section{Some worked out examples}
In this section we present some well-known examples (namely the inflationary Universe, a contracting radiation-dominated Universe and the expanding radiation phase) illustrating the relation between squeezing and $\sigma$. In the inflationary Universe the squeezing parameter $s$ and $\sigma$ are equal and they correctly predict that no $p$ mode survives at horizon re-entry. In the contracting radiation Universe the squeezing parameter seems to point at a similar conclusion.  However, upon more careful inspection it turns out that although half the modes are lost due to squeezing, it is the $y$ mode that is clipped, so that the wrong initial conditions would be injected into the subsequent expanding phase (unless something prevents this at the bounce). This is correctly reflected by the vanishing of  $\sigma$. The analysis of an expanding radiation phase, in turn, merely confirms the results in Section \ref{connector},
within the squeezing formalism.

\subsection{De Sitter inflation}\label{dSinflation}
A particularly simple case is that of de Sitter inflation, for which the general solution is~\cite{Mukhanov:1990me, martin}:
\be\label{solDS}
y(\vk)=\frac{e^{-ik\eta}}{\sqrt{2k}}\left[1-\frac{i}{k\eta}\right]c_0(\vk)+
\frac{e^{ik\eta}}{\sqrt{2k}}\left[1+\frac{i}{k\eta}\right]c_0^\dagger (-\vk).
\ee
For inflation $\omega=k$, and $\eta$ is negative approaching zero at the end of inflation.
This expression can be misleading in that its two terms do not align with the modes $c(\vk,\eta)$ and $c^\dagger (-\vk,\eta)$.
Indeed, working out the associate momentum using (\ref{pexp}), we find:
\be
p(\vk)= i\sqrt{\frac{k}{2}} \left(  e^{i k\eta} c_0^\dagger (-\vk) -e^{-ik\eta} c_0(\vk) \right) \, ,
\ee
so that we can read off from (\ref{eq:yvsc0}) and (\ref{eq:pvsc0}): 
\bea
\sqrt{2k}f_\vk&=&e^{-ik\eta}  \left[1-\frac{i}{k\eta}\right]\\
\sqrt{\frac{2}{k}}g_\vk&=&e^{-ik\eta} \,,
\eea
or from (\ref{uvsfg}) and (\ref{vvsfg}):
\bea
u_\vk &=&e^{-ik\eta}  \left[1-\frac{i}{2k\eta}\right]\\\
v_\vk^\star &=&e^{-ik\eta}  \frac{-i}{2k\eta}.
\eea
Therefore:
\bea
c(\vk,\eta)&=&e^{-ik\eta}  \left[1-\frac{i}{2k\eta}\right]
c_0(\vk)+  
\frac{ie^{ik\eta}  }{2k\eta}
c^\dagger _0(-\vk) \nonumber\\
c^\dagger (-\vk,\eta)&=&
\frac{-ie^{-ik\eta}  }{2k\eta}
c_0(\vk) + 
e^{ik\eta}  \left[1+\frac{i}{2k\eta}\right]
c^\dagger _0(-\vk).\nonumber 
\eea
As we see, the squeezing parameter:
\be
s=|v|^2\rightarrow  \frac{1}{4k^2\eta^2} \rightarrow \infty,
\ee
as $|\eta|\rightarrow 0$, i.e. there is squeezing 
with angles (see also \cite{starob96}): 
\bea
\theta&\rightarrow& \frac {\pi}{2} \\
\phi&\rightarrow& 0
\eea
corresponding to asymptotic relations:
\be 
u=v^\star=-v=-u^\star.
\ee 
Consequently:
\be\label{squeeze-infl}
c(\vk,\eta)\approx c^\dagger (-\vk,\eta)\approx \frac{-i}{2k\eta}(c_0(\vk)-c_0^\dagger(-\vk)).
\ee
In fact, as we can read off from the general solution:
\bea 
y&\approx &\frac{-i}{\sqrt {2}k^{3/2}\eta}(c_0(\vk)-c_0^\dagger(-\vk)),\\
p&\approx &-i \sqrt{\frac{k}{2}}(c_0(\vk)-c_0^\dagger(-\vk)).
\eea
Therefore, 
\be
\sigma\rightarrow \frac{1}{4k^2\eta^2}.
\ee
We see that for inflation the squeezing parameter and the phenomenological parameter $\sigma$ are equal:
\be
s=\sigma
\ee
that is, inflation suppresses the momentum mode as much as it squeezes. In addition the dependence on the initial conditions $c_0(\vk)$ drops out of the final $\sigma$, something which is fact far from true in general, as we saw before.

\subsection{A contracting radiation dominated Universe}\label{contract}
Let us consider a contracting radiation dominated Universe, since this will illustrate many important points to be used later in this paper. 
For radiation $\omega=k/\sqrt{3}$ and  the general solution is~\cite{Mukhanov:1990me, martin}:
\be\label{solrad}
y(\vk)=\frac{e^{-i\omega \eta}}{\sqrt{2\omega}} c_0(\vk)+
\frac{e^{i\omega \eta}}{\sqrt{2\omega }}c_0^\dagger (-\vk) 
\ee
with $\eta$ negative and approaching zero at the Big Crunch.
Again this expression can be misleading because its two terms do not align with the modes $c(\vk,\eta)$ and $c^\dagger (-\vk,\eta)$. Indeed such alignment requires $f_k=g_k/\omega$ and only happens in Minkowski space time. Even though radiation is conformally invariant, the cosmological expansion
cannot be neglected in this respect. From (\ref{pexp}) we see that the momentum conjugate to $y$ is: 
\bea
p(\vk)=-i\sqrt{\frac{\omega}{2}}\left(e^{-i\omega\eta}  \left[1-\frac{i}{\omega\eta}\right] c_0(\vk) \right.\nonumber \\
\left.-
e^{i\omega\eta}  \left[1+\frac{i}{\omega\eta}\right] c_0^\dagger(-\vk)
\right)
\eea
so we can read off from (\ref{eq:yvsc0}) and (\ref{eq:pvsc0}): 
\bea
\sqrt{2 \omega }f_\vk&=&e^{-i\omega\eta} \\
\sqrt{\frac{2}{\omega}}g_\vk&=&e^{-i\omega\eta}  \left[1-\frac{i}{\omega\eta}\right] 
\eea
and this implies:
\bea
u_\vk &=&e^{-i\omega\eta}  \left[1-\frac{i}{2\omega\eta}\right]\\\
v_\vk^\star &=&e^{-i\omega\eta}  \frac{i}{2\omega\eta}.
\eea
As with inflation there is maximal mixing as $|\eta|\rightarrow 0$:
\bea
c(\vk,\eta)&=&e^{-i\omega\eta}  \left[1-\frac{i}{2\omega\eta}\right]
c_0(\vk)+  
\frac{-ie^{i\omega\eta}  }{2\omega\eta}
c^\dagger _0(-\vk) \nonumber\\
c^\dagger (-\vk,\eta)&=&
\frac{ie^{-i\omega\eta}  }{2\omega\eta}
c_0(\vk) + 
e^{i\omega\eta}  \left[1+\frac{i}{2\omega\eta}\right]
c^\dagger _0(-\vk),\nonumber\\\label{ccdagrad}
\eea
and indeed the squeezing parameter is precisely the same as inflation's:
\be
s=|v|^2= \frac{1}{4\omega^2\eta^2}.
\ee
However, the squeezing angles are different:
\bea
\theta&\rightarrow& \frac {\pi}{2} \\
\phi&\rightarrow& \frac{\pi}{2},
\eea
reflecting the different asymptotic relations:
\be
u=-v^\star=v=-u^\star\; .
\ee
Instead of (\ref{squeeze-infl}) we have:
\be\label{squeeze-rad}
c(\vk,\eta)=-c^\dagger (-\vk,\eta)=\frac{-i}{2\omega \eta}(c_0(\vk)+c_0^\dagger(-\vk)).
\ee
but  squeezing now implies: 
\bea 
y&\approx&\frac{1}{\sqrt {2\omega } }(c_0(\vk)+c_0^\dagger(-\vk))\\
p&\approx&\frac{-1}{\sqrt {2\omega } \eta}(c_0(\vk)+c_0^\dagger(-\vk))\, .
\eea
We still loose half the degrees of freedom, but they happen to be the wrong ones. 
This is reflected in 
\be
\sigma\approx \frac{\omega^2\eta^2}{4}\rightarrow 0,
\ee
valid for all initial conditions. 
This example shows that squeezing is not enough: the squeezing angles matter too. In this case $s\gg 1$ but $\sigma \ll 1$.

\subsection{Radiation dominated re-entry}\label{rad-reentry}
As the previous example shows the squeezing parameter $s$ may be very unfit for the purpose of measuring the required reduction in phase space. But there is another situation in which this is even more blatant: 
an {\it expanding} radiation dominated universe which receives a spectrum of fluctuations from a previous phase. 
Then, the algebra is the same as that of the previous subsection, and so we would expect 
$s\approx 1/(4\omega^2\eta)^2\rightarrow 0$ but $\sigma\approx \omega^2\eta^2/
4\rightarrow\infty$ (since now $\eta>0$),  that is unsqueezing and suppression of momentum. From Section~\ref{connector} we know that the latter is true. The former, however, is wrong, as we now show. 

Although the algebra leading to (\ref{ccdagrad}) is applicable to this case, the interpretation and labelling are incorrect. Taking the $\eta\rightarrow \infty$ limit, we see that the coefficients $c_0(\vk)$ and $c_0^\dagger (-\vk)$ in that expression are in fact the values of  $c(\vk,\eta)$ and $c^\dagger (-\vk,\eta)$ as $\eta\rightarrow \infty$, and so we change their label from zero to infinity throughout, namely in (\ref{ccdagrad}):
\bea
c(\vk,\eta)&=&e^{-i\omega\eta}  \left[1-\frac{i}{2\omega\eta}\right]
c_\infty(\vk)+  
\frac{-ie^{i\omega\eta}  }{2\omega\eta}
c^\dagger _\infty(-\vk) \nonumber\\
c^\dagger (-\vk,\eta)&=&
\frac{ie^{-i\omega\eta}  }{2\omega\eta}
c_\infty(\vk) + 
e^{i\omega\eta}  \left[1+\frac{i}{2\omega\eta}\right]
c^\dagger _\infty(-\vk).\nonumber 
\eea
Expanding these expressions for $\omega\eta\ll 1$ to find $c_0(\vk)$ and $c_0(-\vk)$, we obtain to order $\omega\eta_0$,  after some reorganization: 
\bea
c_0(\vk)+c^\dagger_0(-\vk,)&=&c_\infty(\vk)+c_\infty^\dagger(-\vk)\nonumber\\
&&-i\omega\eta_0 (c_\infty(\vk)-c_\infty^\dagger(-\vk))\\
c_0(\vk)-c_0^\dagger(-\vk)&=&\frac{-i}{\omega\eta_0}(c_\infty(\vk)+c_\infty^\dagger(-\vk))\label{ctemp2}.
\eea
(In (\ref{ctemp2}) we dropped a term in $\omega \eta_0$ in the coefficient of $c_\infty(\vk)+c_\infty^\dagger(-\vk)$, since this is negligible compared to the other term; also the coefficient of $c_\infty(\vk)-c_\infty^\dagger(-\vk)$ in the second expression is zero to order $\omega\eta_0$)). This can be inverted as:
\bea
c_\infty(\vk)+c_\infty^\dagger(-\vk)&=&i\omega\eta_0(c_0(\vk)-c_0^\dagger(-\vk))\\
c_\infty(\vk)-c_\infty^\dagger(-\vk)&=&\frac{i}{\omega\eta_0}(c_0(\vk)+c_0^\dagger(-\vk))\nonumber\\
&& + (c_0(\vk)-c_0^\dagger(-\vk)).
\eea
We see that even if the $c_0(\vk)$ and $c_0(-\vk)$ are unconstrained we are led at late times to a squeezed mode with $c_\infty(\vk)\approx -c^\dagger_\infty(-\vk)$. The only way this can be avoided is by setting $c_0(\vk)+c_0^\dagger (-\vk)\ll c_0(\vk)-c_0^\dagger (-\vk)$. Otherwise
\bea
c_\infty(\vk)+c_\infty^\dagger (-\vk)&=& i \omega\eta_0(c_0(\vk)-c_0^\dagger (-\vk))\label{squeeze-central}\\
c_\infty(\vk)-c_\infty^\dagger (-\vk)&=&\frac{i}{\omega\eta_0}(c_0(\vk)+c_0^\dagger (-\vk))
\eea
and these can be reorganized as 
\be
\frac{|c_\infty(\vk)-c_\infty^\dagger (-\vk)|}{|c_\infty(\vk)+c_\infty^\dagger (-\vk)|}=
\frac{1}{(\omega\eta_0)^2} \frac{|c_0(\vk)+c_0^\dagger (-\vk)|}{|c_0(\vk)-c_0^\dagger (-\vk)|}.
\ee
This is just (\ref{Sigma-sigma}) derived using the squeezing framework. 

We see that ``squeezing'' does take place in the setting of an expanding universe, in spite of the negative forecast by parameter $s$. Even if the $c_0(\vk)$ and $c_0(-\vk)$ are unconstrained we are led at late times to a squeezed mode with $c_\infty(\vk)\approx -c^\dagger_\infty(-\vk)$. A standing wave (with the correct temporal phase) is then produced. In this sense inflation is overkill, in that it feeds a squeezed state with 
$c_0(\vk)\approx c_0^\dagger(-\vk)$ in the radiation epoch. This is not needed. All that we need is that 
$c_0(\vk)\approx - c_0^\dagger(-\vk)$ does {\it not} happen. 

A case where this exception happens is at the end of a contracting phase. If we consider a collapsing radiation dominated universe followed by a bounce incapable of filtering out the growing mode, then we do feed into the expanding phase precisely the squeezed mode that might be problematic (that with $c_0(\vk)\approx - c_0^\dagger(-\vk)$). Then a travelling wave in due time might enter the horizon, contradicting phenomenology. In fact we know that this has to be the case, given the time symmetry of the evolution, and that we assumed an unsqueezed set of traveling waves exited the horizon in the contracting phase. In practice this can be avoided by filtering out the pathological mode at the bounce, something that is required anyway to produce the correct power spectrum~\cite{Finelli:2001sr}.

\section{Bimetric VSL theories}\label{VSL}
We now examine the status of the various matters discussed so far in alternative theories of the Early Universe,
starting with bimetric VSL theories. We refer the reader to the literature for background and details~\cite{Jprl,Jprd,Fedo,Ny,JN}. In these theories there are two metrics and these define two frames (an Einstein and a matter frame). 

In the Einstein frame the second order action for the fluctuations is that of (anti-)DBI theories, or more generally,
of theories with a varying speed of sound:
\bea
S_2&=&\int d^3 k d\eta \, z^2[\zeta'^2 +c_s^2 k^2 \zeta^2]
\eea
with $z=\frac{a}{c_s}$, with $c_s$ to be computed as explained in~\cite{Jprd}. 

A dual frame, where the speed of sound/light is constant, may be obtained by 
defining a new time: 
\be
\tau=\int c_s d\eta.
\ee
In the dual frame Einstein's gravity is no longer valid, even at the zeroth order. Under the conditions necessary to solve the horizon problem in the Einstein frame (viz. that  $c_sk\eta$ drops in time sufficiently fast) we find that $\tau$ runs from $-\infty$ to zero. Therefore, we kinematically have inflation in the dual frame, although dynamically this is due to a very modified theory of gravity, rather than an inflaton (see~\cite{Fedo} for more details).   

In the dual frame the action is~\cite{Fedo}: 
\bea
S_2&=&\int d^3 k d\tau\, q^2[\dot\zeta^2 +k^2 \zeta^2]\\
q&=&\frac{a}{\sqrt{c_s}}
\eea
(here a prime denotes derivative with respect to $\eta$, a dot to $\tau$), and the calculations simplify significantly. 
The most general scaling solutions for these theories are labeled by two constant parameters~\cite{Fedo}:
\bea
\epsilon&=&-\frac{\dot H}{H^2}\\
\epsilon_s&=&\frac{\dot c_s}{c_sH}\, .
\eea
They are the power-law solutions:
\bea
a&\propto (-\tau)^\frac{1}{\epsilon 
_s+\epsilon -1}\\
c_s&\propto (-\tau)^\frac{\epsilon_s}{\epsilon 
_s+\epsilon -1}.
\eea
It is then not difficult to find the conditions for scale invariance of vacuum fluctuations. It must be
$\epsilon_s=-2\epsilon$, since then  
\be
q\propto \frac{1}{-\tau}.
\ee
Under these circumstances the kinematics of the fluctuations is in every way identical to that of the fluctuations in de Sitter inflation
(and thus, their scale-invariance, but see~\cite{Jprl} for a more complete derivation). This is true even though the dynamics is very different. 
For example we do not have de Sitter expansion in the dual frame. The two facts can be reconciled by realizing that 
gravity is modified in the dual frame (precisely because it is Einstein in the original varying-$c$ frame).

We can now compute the squeezing parameter and $\sigma$ for these theories. We have:
\be\label{solBIM}
\tilde y(\vk)=\frac{e^{-ik\tau}}{\sqrt{2k}}\left[1-\frac{i}{k\tau}\right]\tilde c_0(\vk)+
\frac{e^{ik\tau}}{\sqrt{2k}}\left[1+\frac{i}{k\tau}\right]\tilde c_0^\dagger (-\vk)
\ee
and the rest of the algebra of subsection~\ref{dSinflation} follows through (e.g. regarding $\tilde p$, the $\tilde c$, etc) leading to 
\be
\tilde s\approx \tilde \sigma \approx \frac{1}{4k^2\tau^2}
\ee
It is straightforward to show that these conclusions are invariant under frame transformation. The 
canonical pair in the Einstein frame:
\bea
y&=&z \zeta \\
p&=&z\zeta'=y'-\frac{z'}{z}y
\eea
can be compared with its counterpart in the dual frame
\bea
\tilde y&=&q \zeta \\
\tilde p&=&q \dot \zeta=\dot {\tilde y}-\frac{\dot q}{q}\tilde y.
\eea
to give $\tilde y=y\sqrt{c_s}$ and $\tilde p=p/\sqrt{c_s}$.
Comparing (\ref{eq:yvsc}) and (\ref{eq:pvsc}) (valid in the Einstein frame) with 
\bea
\tilde y(\vk,\tau)&=&\frac{\tilde c(\vk,\tau)+\tilde c^\dagger (-\vk,\tau)}{\sqrt{2 k}} \label{eq:yvsc1}\\
\tilde p(\vk,\tau)&=&-i \sqrt{\frac{ k}{2}}(\tilde c(\vk,\tau)- \tilde c^\dagger (-\vk,\tau))\label{eq:pvsc1}
\eea
valid in the dual frame, we see that:
\bea
c(\vk,\eta)&=&\tilde c(\vk,\tau)\\
c^\dagger (-\vk,\eta)&=&\tilde c^\dagger (-\vk,\tau)\, 
\eea
and so $\tilde s=s$. In addition
\be
\tilde \sigma=\frac{|\tilde y|^2 k^2}{4 |\tilde p|^2}= \frac{|y|^2 \omega^2}{4 |p|^2}=\sigma
\ee
It is therefore trivial to transform the solutions and conclusions from the dual frame to the Einstein frame. 
We conclude that scale invariant bimetric VSL is very similar to inflation with regards to squeezing. 

Although we have used the vacuum scale-invariant solution to make our point, the calculation can be repeated for near  scale-invariant solutions, as well as those where the fluctuations are of a thermal origin, with similar conclusions.  For example, for the latter the solution is approximately~\cite{Ny,JN}: 
\be
\tilde y=\frac{\sqrt{-\pi\tau}}{2}(H^{(1)}_1(-k\tau)c_0(\vk) + H^{(2)}_1(-k\tau)c^\dagger_0(-\vk) )
\ee
The final conclusions are qualitatively the same, even though the algebra is more complicated. 

The vacuum scale-invariant solution is simpler because it invokes Hankel functions of order $\nu=3/2$. But for more complex situations the story is invariably the same. The modes leave the horizon as Hankel functions (travelling waves). Outside the horizon (either in the primordial phase, or later in the standard radiation dominated phase) these are split as $H_\nu^1=J_\nu+i Y_\nu$ and $H_\nu^2=J_\nu-i Y_\nu$, i.e. as 
Bessel functions $J$ (regular at the origin; the growing mode) and $Y$ (divergent at the origin; the decaying mode). Natural selection spits out $J$ instead of Hankel functions at horizon re-entry. This amounts to a standing wave with the correct temporal phase.

This comment is also pertinent for modes that reenter the horizon after the end of the radiation dominated phase (see the last paragraph of Section~\ref{connector}). Then our earlier arguments cannot be made with sine and cosine waves and simple travelling waves (corresponding to Bessel/Hankel functions of order 1/2). However the appropriate real $J_\nu$ mode will play the role of the standing wave with the correct temporal phase, no longer a simple sine wave.

\section{Theories with modified dispersion relations}~\label{DSR}

It is also interesting to study the evolution of perturbations in theories where they obey dispersion relations that are modified at some ultraviolet (UV) scale - usually assumed to be the Planck scale. In fact, such theories with modified dispersion relations (MDR) can generate a scale invariant spectrum of primordial perturbations \cite{Amelino-Camelia:2013tla, Amelino-Camelia:2013wha, Amelino-Camelia:2013gna, Amelino-Camelia:2015dqa} without invoking a primordial phase of exponential expansion such as inflation.  This was shown explicitly by assuming that the UV form of the dispersion relation is the one that appears in Horava-Lifshitz theory \cite{Horava}, so that  the velocity $c_{\gamma}=\frac{d E}{d p}$ depends on time and wavelength $k$ via
\be
c_{\gamma}= \left(\frac{\lambda k}{a(\eta)}\right)^{\gamma}\,,\label{eq:speedgamma}
\ee
where $\lambda^{-1}$ is the UV scale and $\gamma$ is a dimensionless parameter.
Starting from the second order Einstein-Hilbert action for primordial perturbations, one finds that the evolution equation takes the form:
\be\label{yeq}
y(\vk,\eta)''+ \left[ \left(\frac{\lambda k}{a(\eta)}\right)^{2 \gamma} k^{2} -\frac{a(\eta)''}{a(\eta)}\right]y(\vk,\eta)=0\
\ee
(which is just the usual $y$ equation with $c$ given by (\ref{eq:speedgamma}) and $z=a$).
The normalized vacuum WKB solution for modes inside the horizon, when the first term in brackets dominates, is, up to a phase,
\be
y \sim  \frac{a(\eta)^{\gamma/2}}{(\lambda k)^{\gamma/2}\sqrt{ k}}\,.
\ee
This is clearly scale invariant when $\gamma=2$, and since $y(\vk,\eta) \sim a$ outside the horizon, by matching the two solutions it can be easily seen that scale invariance is preserved once the modes exit the horizon, independently of the background equation of state. 

\subsection{The critical, scale-invariant model}
In this subsection we study in detail the evolution of $y$ and its conjugate momentum $p$ in the MDR model that produces scale invariant perturbations (the one with $\gamma=2$ in Eq.~\eqref{eq:speedgamma}). Because scale invariance in achieved independently of the background evolution, we will keep the equation of state general,  so that 
\bea
a(\eta)&\propto &\eta^m\\
m&=&\frac{2}{1+3w}\,.
\end{eqnarray}
Slightly smaller values of $\gamma$ lead to the red spectrum of perturbations that is observed in the CMB. We review this case in the next subsection.

The dynamical equation (\ref{yeq}) then becomes:
\be
y(\vk,\eta)''+ \left[\left(\frac{\lambda k}{\eta^{m}}\right)^4 k^{2} -2 \frac{1-3w}{(1+3w)^{2}}\eta^{-2}\right]y(\vk,\eta)=0\,. \label{MDReom}
\ee
Solving the horizon problem with $\eta>0$ is possible if $-\frac{1}{3} < w < 1$, so we will restrict the allowed values of $w$ to this range. Then, setting $\alpha=2m-1>0$, the full solution to \eqref{MDReom} is 
\bea
y&=&\frac{1}{\sqrt{2 \omega}}\left[e^{\frac{i k^{3}\lambda^{2}}{\alpha \eta^\alpha}}
c_{0}(\vk)+
e^{\frac{-i k^{3}\lambda^{2}}{\alpha \eta^\alpha}}c_{0}^{\dagger}(-\vk)\right]\ .\label{yMDR}
\eea
Note that the signs of the exponentials have been correctly swapped, since $\vk$ labels the direction of motion of the propagating wave, and the temporal phase $\omega\eta$ is a decreasing function of time. 
With the signs used above, upon multiplication by $\exp(i\vk\cdot \vx)$ (cf. Eq.~\ref{FT1}) the first mode moves along $\vk$, the second along $-\vk$. 
Note also that for a generic  $\gamma$ there is no solution in terms of elementary functions (see~\cite{Magueijo:2008yk}). This is only possible for $\gamma=2$ because the solution is given in terms of Hankel functions $H_{1/2}$ and $H_{-1/2}$, which are in fact the simple trigonometric functions encoded in \eqref{yMDR}.

We can now compute the relevant quantities for our analysis. 
The conjugate variable $p=y'-\frac{a'}{a}y$ is given by:
\bea
p&=&- i\sqrt{\frac{\omega}{2}}
\left[e^{\frac{i k^{3}\lambda^{2}}{\alpha \eta^\alpha}}
c_{0}(\vk)-
e^{\frac{-i k^{3}\lambda^{2}}{\alpha \eta^\alpha}}c_{0}^{\dagger}(-\vk)\right]\,.
\label{pMDR}
\eea
At late times one has
\bea
y(\vk,\eta)&\approx&\frac{1}{\sqrt{2\omega}} [c_{0}(\vk)+ c_{0}^{\dagger}(-\vk)]\\
p(\vk,\eta)&\approx& -i\sqrt{\frac{\omega}{2}} \left[  c_{0}(\vk)-  c_{0}^{\dagger}(-\vk) \right]\,,
\eea
so that 
\be
\sigma \sim 1.
\ee
Therefore, the scale invariant MDR model is a case where the requirements demanded by phenomenological constraints are met without overkill. The model also causes no squeezing. 
Indeed 
\be
s= 0\,,
\ee
as can be computed from  \eqref{yMDR} and \eqref{pMDR} using the formulas in Section~\ref{squeeze}.

\subsection{Models with a slightly red spectrum}\label{DSRred}
For models departing from exact scale-invariance the calculations are more intricate (and involve Bessel functions, just like for bimetric VSL scenarios in this regime). However a significant simplification is made possible by realizing that we can focus on the outside the horizon regime in order to study the status of ``squeezing''. In this regime we have
\be
y'' \approx \frac{a''}{a} y
\ee
so that for any power-law $a$ the general solution takes the form:
\be
y=Aa +B\frac{\eta}{a} .
\ee
The first term is the growing mode, the second the decaying mode. At late times $y\approx A a$, but since this mode does not generate any momentum, we must look at the decaying mode to find
\be
p=a{\left(\frac{y}{a}\right)}'\approx \frac{C}{a}
\ee
with $C\propto B$. Therefore in any model of this type
\be\label{sigmaDSR}
\sigma=\frac{\omega^2 |y|^2}{|p^2|}=k^{2}\left(\frac{\lambda k}{a}\right)^{2\gamma}\frac{A^2 a^2}{C^2/a^2}\propto
a^{4-2\gamma}.
\ee
We confirm that $\sigma$ is constant for $\gamma=2$, but could grow or decay with expansion if $\gamma\neq 2$. 

At this point we could try to input into our calculation the fact that the observed primordial spectrum is slightly red ($n_S\approx 0.96$) and this could be due to an anomalous value of $\gamma<2$, as suggested in~\cite{Amelino-Camelia:2013tla}. This would correspond to a fractional UV dimensionality of spacetime slightly above 2 (the exact value depending on the equation of state). If such an anomalous $\gamma$ led to the growth of the momentum mode (decay of $\sigma$) this could turn MDR into something similar to the bouncing universe. Eq.~\eqref{sigmaDSR} shows that this only happens for blue spectra (with $\gamma>2$ and $n_s>1$); for red spectra $\sigma$ actually grows during the phase where the MDRs are active.

\section{Conclusions}
To conclude, in this paper we considered the status of ``squeezing'' in scenarios alternative to inflation, to assess whether its blessings are peculiar to inflation. We found just the opposite. Firstly, we noted that from a purely phenomenological standpoint, all that matters regarding the effects of squeezing is that at late times (at horizon re-entry) the fluctuations form standing waves with the correct temporal phase (specifically, a sine wave). Admixture of
the complementary standing wave mode (the cosine wave) would have an effect similar to the addition of an isocurvature mode, leading to out of phase Doppler peaks, or a softening of the oscillatory structure. 

Bearing this in mind we were then able to conclude that just about any model complies with the phenomenological requirements. More concretely, we found that it is enough for the primordial phase producing the fluctuations to feed into the expanding radiation dominated epoch a configuration for which the conjugate momentum of the fluctuations is not abnormally high. Inflationary squeezing in fact heavily suppresses  the momentum mode, but this is surplus to requirement. A similar overkill takes place in bimetric VSL scenarios,  as well a models based on MDRs producing a red spectrum. But in fact inputing into the standard radiation phase a momentum mode of the same amplitude as a momentum-free mode is sufficient, and this happens in scale-invariant scenarios based on MDRs. String gas cosmologies \cite{Nayeri:2005ck} are also possibly in this category, although we have not investigated the matter in detail.

In view of our findings, the only problematic scenario among those we have studied would be a symmetric bouncing universe model, for which the bounce is incapable of filtering out the momentum mode (which is the growing mode in the collapsing phase). In such a model a very large momentum mode is fed into the expanding radiation phase, so that when the modes re-enter the horizon they form travelling, rather than standing waves. But such a model is known to be pathological for all sorts of other reasons \cite{Finelli:2001sr}, and filtering out the momentum mode is a basic requirement to be imposed upon the bounce for any viable model. 

One may wonder about the origin of the resilience of models with regards to the production of standing waves with the correct temporal phase. Standing waves are the result of the imposition of a constraint upon travelling waves. This constraint is enforced in {\it any} late time expanding universe because while the modes are outside the horizon  there is a growing mode (the momentum-free mode with with $\zeta\neq 0$ and $\zeta'=0$) and a decaying mode (the momentum mode with $\zeta\neq 0$ and $\zeta'=0$). As the name indicates, the  latter decays away, leaving a constraint between the two travelling waves when they re-enter the horizon. This amounts to the production of a standing wave with a sine phase in time. Squeezing and momentum suppression in inflation and other scenarios start doing this job, but in fact this is not needed.  The only way to produce a travelling wave at late times  would be to feed into the radiation epoch an extremely large momentum mode, which would then decay until it became of the same amplitude as the momentum-free mode when the two modes re-entered the horizon. If the decaying mode were given an even higher initial amplitude, a cosine standing wave would be produced. The difficulty in conjuring up such an initial condition for the expanding radiation dominated phase explains why it is so easy to satisfy the observational constraints.

We have made our point for scalar fluctuations, but of course they apply equally well to gravity waves, should the model produce them.  In work in preparation we show how novelties arise in this set up regarding MDR scenarios capable of producing a blue spectrum of primordial gravitational waves~\cite{prep2}.

\section*{Acknowledgments}
We thank Robert Brandenberger, Carlo Contaldi and Marco Peloso for discussions that led to this paper.  We acknowledge support from the John Templeton Foundation. JM was also supported by  an STFC consolidated  grant.

%=====================================================================
%=====================================================================
%=====================================================================

\end{document}